\documentclass[12pt]{article}

\usepackage{geometry}
\usepackage{lmodern}
\usepackage{amssymb,amsmath}
\usepackage{ifxetex,ifluatex}
\usepackage{hyperref}
\usepackage{parskip}
\usepackage[multiple]{footmisc}
\usepackage{longtable,booktabs}
\usepackage{graphicx}
\def\fps@figure{htbp}
\providecommand{\tightlist}{%
  \setlength{\itemsep}{0pt}\setlength{\parskip}{0pt}}
\usepackage{authblk}

\title{quTARANG: A python GPE solver to study turbulence in quantum systems}
\date{}

\author[1]{Shawan Kumar Jha\footnote{\small{These authors contributed equally to this work.}}$^,$\footnote{\small{Correspondence: shawankumar@iitg.ac.in}}$^,$}

\author[2]{Sachin Singh Rawat$^{*,}$}
\author[2]{Mahendra Kumar Verma} 
\author[1]{Pankaj Kumar Mishra}
\affil[1]{Department of Physics, Indian Institute of Technology Guwahati, Guwahati 781039, India}
\affil[2]{Department of Physics, Indian Institute of Technology Kanpur, Kanpur 208016, India}

\begin{document}
\maketitle
\begin{abstract}
quTARANG is a Python-based general-purpose Gross-Pitaevskii Equation (GPE) solver. It can solve GPE in 1D, 2D and 3D and has the ability to run on both CPU and GPU. It has been developed to study turbulence in quantum systems, specifically in atomic Bose-Einstein condensates, and can be used to study   different quantities, such as the varied spectra associated with quantum turbulence.
\end{abstract}

\section{Summary}\label{summary}

Turbulence is a phenomenon associated with the chaotic nature of flows
in space and time solely arising due to the nonlinear nature of the
interactions. Turbulence in classical fluids, as best characterised by
the Navier-Stokes equation, remains unresolved to this day. The complex
nonlinear interactions at various length and time scales make it a
difficult problem to handle analytically as well as numerically. One
recent approach to shed some light on this long-standing problem has
been the study of turbulence in quantum fluids. The zero viscosity and
quantized vortices of quantum fluid systems like Bose-Einstein
Condensates (BECs) \cite{madeira2020quantum} distinguish it from its
classical counterparts. BEC is a state of matter where Bose particles
occupy the ground state upon cooling to a very low temperature and thus
can be represented by a macroscopic wave function. One can model the
dynamics of BECs using the mean-field Gross-Pitaevskii Equation (GPE)
given by \cite{muruganandam2009fortran} \begin{equation}\label{eqn:GPE}
i\hbar\partial_t\psi(\vec{r},t) = -\frac{\hbar^2}{2m}\nabla^2\psi(\vec{r},t) + V(\vec{r},t)\psi(\vec{r},t) + NU_0|\psi(\vec{r},t)|^2\psi(\vec{r},t),
\end{equation}

where \(\psi(\vec{r},t)\) is the macroscopic complex wave function,
\(m\) is the atomic mass, \(V(\vec{r},t)\) is the trapping potential,
\(N\) is the number of particles,
\(\displaystyle U_0=(4\pi\hslash^2a_s)/m\) is the nonlinear interaction
parameter and \(a_s\) denotes the scattering length for the interaction
of the atomic particles.

Our quantum simulator code \texttt{quTARANG}\footnote{\href{https://github.com/sachinrawat2207/quTARANG}{github.com/sachinrawat2207/quTARANG}} is primarily designed for
studying quantum turbulence in BECs by solving the GPE in laminar as
well as in the turbulent regime.

\section{Statement of Need}\label{statement-of-need}

\texttt{quTARANG} is a robust and easy-to-use application that solves
the GPE with Graphics Processing Unit (GPU). GPUs are specialized units
primarily designed for image processing and for performing massive
multigrid simulations at high speed. They have a large number of
parallelizing units compared to CPUs, due to which they are being widely
used to speed up code. There are no packages available in python that
can solve turbulent GPE in 2D and 3D on both CPUs and GPUs. There exist,
however, some software packages in other languages that can solve the
GPE such as GPELab \cite{Antoine2014}, Massively Parallel Trotter-Suzuki
Solver \cite{Wittek2013}, CUDA-enabled GPUE \cite{schloss2018gpue}, a
split-step Crank-Nicolson based Fortran code
\cite{muruganandam2009fortran} and MPI-OpenMP enabled Gross-Pitaevskii
Solver (GPS) \cite{kobayashi2021quantum}.

\begin{figure}
\centering
\includegraphics[scale = 0.6]{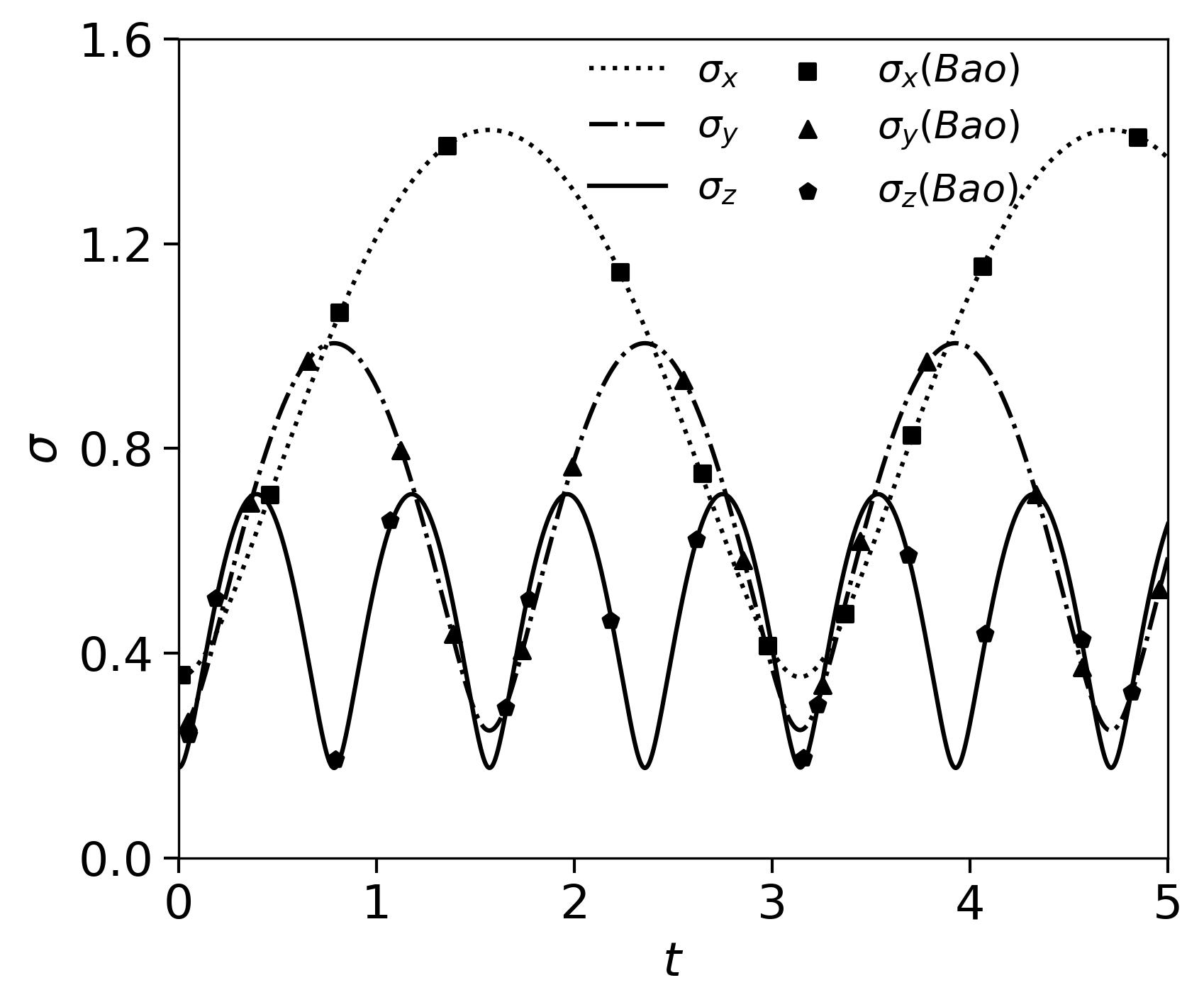}
\caption{A comparison of the dynamical evolution of the root-mean-square
size of the condensate in the \(x\) (\(\sigma_x\)), \(y\)
(\(\sigma_y\)), and \(z\) (\(\sigma_z\)) direction obtained from our
simulation (for 3D GPE) and those obtained by Bao et
al. \cite{bao2003numerical}. A perfect match is being
obtained.\label{fig:dynamic}}
\end{figure}

\section{Numerical Scheme and
functionalities}\label{numerical-scheme-and-functionalities}

We have chosen the frequency (\(\omega^{-1}\)) as the time scale,
oscillator length \(a_0 = \sqrt{\hbar/m\omega}\) as the characteristic
length scale and the harmonic oscillator ground state energy
\(\hbar\omega\) as the energy scale. With this formalism, the
non-dimensional variables can be written as \(t'=\omega t\),
\(\vec{r}'=\vec{r}/a_0\) and \(\psi' = a_0^{3/2}\psi\). In what follows,
we omit the prime \((')\) from the variables. The non-dimensional form
of the GPE is given by \cite{muruganandam2009fortran}

\begin{equation}\label{eqn:ndGPE}
i\partial_t\psi(\vec{r},t)= -\frac{1}{2}\nabla^2\psi(\vec{r},t) + V(\vec{r},t)\psi(\vec{r},t) +g|\psi(\vec{r},t)|^2\psi(\vec{r},t)
\end{equation} where \(g\) is the non-dimensional interaction parameter.

\texttt{quTARANG} uses a pseudo-spectral scheme, Time-splitting spectral
(TSSP) method \cite{bao2003numerical}, to solve the dynamics of the GPE.
The main advantage of using the TSSP scheme is that it is
unconditionally stable and conserves the total particle number.

The ground state calculations in \texttt{quTARANG} are done by using
TSSP with an imaginary time propagation method. In this method, one
replaces \(t\) with \(-it\). As we propagate in imaginary time, the
eigenstates with higher energies begin to decay faster than the ground
state as a result of which only the ground state survives. The
wavefunction needs to be normalised at each time step in order to
conserve total particle number.

\texttt{quTARANG} is equipped with various features, which include:

\begin{enumerate}
\def\labelenumi{\arabic{enumi}.}
\tightlist
\item
  Ground state calculations for different potentials such as harmonic
  and anharmonic trap, optical lattice potential, time-dependent
  potential and stochastic potential.
\item
  Long-time dynamical evolution of different states using either CPUs or
  GPUs.
\item
  Computation of different quantities relevant to the study of
  turbulence phenomenon in BECs, such as components of kinetic energy
  (KE) and various spectra (compressible KE spectrum, incompressible
  KE spectrum and particle number spectrum).
\end{enumerate}

\begin{longtable}[]{@{}cccccc@{}}
\caption{The condensate width (\(r_{rms}\)) and chemical potential
(\(\mu\)) obtained for the ground state using \texttt{quTARANG}.
\(r^*_{rms}\) and \(\mu^*\) are the corresponding values from
\cite{muruganandam2009fortran} for comparison.
\label{table:gstate}}\tabularnewline
\toprule
\(Dimension\) & \textbf{\(g\)} & \textbf{\(r_{rms}\)} &
\textbf{\(r^*_{rms}\)} & \textbf{\(\mu\)} &
\textbf{\(\mu^*\)}\tabularnewline
\midrule
\endfirsthead
\toprule
\(Dimension\) & \textbf{\(g\)} & \textbf{\(r_{rms}\)} &
\textbf{\(r^*_{rms}\)} & \textbf{\(\mu\)} &
\textbf{\(\mu^*\)}\tabularnewline
\midrule
\endhead
& -2.5097 & 0.87771 & 0.87759 & 0.49987 & 0.49978\tabularnewline
& 0 & 1.00000 & 1.00000 & 1.00000 & 1.00000\tabularnewline
& 3.1371 & 1.10504 & 1.10513 & 1.42009 & 1.42005\tabularnewline
2 & 12.5484 & 1.30656 & 1.30687 & 2.25609 & 2.25583\tabularnewline
& 62.742 & 1.78722 & 1.78817 & 4.61136 & 4.60982\tabularnewline
& 313.71 & 2.60122 & 2.60441 & 10.07639 & 10.06825\tabularnewline
& 627.42 & 3.07914 & 3.08453 & 14.20569 & 14.18922\tabularnewline
& & & & &\tabularnewline
& 0 & 1 & 1 & 3.0000 & 3.0000\tabularnewline
& 18.81 & 1.3778 & 1.3249 & 4.3618 & 4.3611\tabularnewline
& 94.05 & 1.8222 & 1.7742 & 6.6824 & 6.6797\tabularnewline
3 & 188.1 & 2.0881 & 2.0411 & 8.3718 & 8.3671\tabularnewline
& 940.5 & 2.8912 & 2.8424 & 14.9663 & 14.9487\tabularnewline
& 1881 & 3.3268 & 3.2758 & 19.5058 & 19.4751\tabularnewline
& 7524 & 4.3968 & 4.3408 & 33.5623 & 33.4677\tabularnewline
& 15048 & 5.0497 & 4.9922 & 44.1894 & 44.0234\tabularnewline
\bottomrule
\end{longtable}

\section{Results}\label{results}

We have calculated the ground state and dynamics for given sets of
initial conditions and compared them with the standard results.

\begin{enumerate}
\def\labelenumi{\arabic{enumi}.}
\item
  \textbf{Validation of ground state} : The initial condition for 2D and
  3D cases are given as:

  2D :
  \(\psi(\vec{r},0) = \left(\frac{1}{\pi}\right)^{1/2}e^{-\frac{(x^2+y^2)}{2}}\),
  ~\(V(\vec{r}) = \frac{1}{2}(x^2 + y^2)\)

  3D :
  \(\psi(\vec{r},0) = \left(\frac{1}{\pi}\right)^{3/4}e^{-\frac{(x^2+y^2+z^2)}{2}}\),
  ~\(V(\vec{r}) = \frac{1}{2}(x^2 + y^2 + 4z^2)\)

  The ground state obtained by \texttt{quTARANG} has been compared with
  that obtained by using the finite difference code of Muruganandam and
  Adhikari \cite{muruganandam2009fortran}. The results for harmonic
  potential well for 2D and 3D are in good agreement with Muruganandam
  and Adhikari \cite{muruganandam2009fortran} as shown in
  Table \ref{table:gstate}. The root-mean-square size (\(r_{rms}\)) of the
  condensate is defined as
  \(r_{rms} = \left(\int r^2 |\psi(\vec{r}, t)|^2dV \right)^{1/2}\).
\item
  \textbf{Validation of dynamics} : We have validated the dynamic
  evolution of a state by comparing our results with Bao et
  al. \cite{bao2003numerical} for the following condition:
  \[\psi(\vec{r},0)=\frac{(\gamma_y\gamma_z)^{1/4}}{\sqrt{(\pi\epsilon_1)^{3/4}}}e^{-\frac{(x^2+\gamma_yy^2+\gamma_zz^2)}{2\epsilon_1}}, \ \ \ V(\vec{r},0)=\frac{1}{2}(x^2+\gamma_y^2y^2+\gamma_z^2z^2),\]
  where \(\gamma_y = 2.0\), \(\gamma_z = 4.0\), \(\epsilon_1 = 0.25\)
  and \(g = 0.1\). Fig \ref{fig:dynamic} shows the comparisons of the
  rms size of the condensate in \(x\), \(y\) and \(z\) directions at
  different times calculated by using our code and those obtained by Bao
  et al. \cite{bao2003numerical}. The results obtained from the
  \texttt{quTARANG} are in good agreement with the results obtained from
  Bao et al.~for the same initial conditions.
\end{enumerate}


\end{document}